\documentclass[aps,showpacs,superscriptaddress]{revtex4}

\usepackage{amssymb,amsfonts}

\usepackage[dvips]{graphicx}

\def\go{\rightarrow}

  \def\d{\delta } \def\D{\Delta }

\def\s{\sigma }

\def\s1{{\textbf{1}}}
\def\half{{1\over 2}}

\begin{document}

\title[Social Influencing and Associated Random Walk Models]
{Social Influencing and Associated Random Walk Models: Asymptotic
Consensus Times on the Complete Graph}

\author{W. Zhang}
\affiliation{Social and Cognitive Networks Academic Research Center,
Rensselaer Polytechnic Institute, 110 8$^{th}$ Street, Troy, NY 12180--3590, USA}
\affiliation{Department of Mathematics,
Rensselaer Polytechnic Institute, 110 8$^{th}$ Street, Troy, NY 12180--3590, USA}

\author{C. Lim}
\affiliation{Social and Cognitive Networks Academic Research Center,
Rensselaer Polytechnic Institute, 110 8$^{th}$ Street, Troy, NY 12180--3590, USA}
\affiliation{Department of Mathematics, Rensselaer Polytechnic Institute,
110 8$^{th}$ Street, Troy, NY 12180--3590, USA}

\author{S. Sreenivasan}
\affiliation{Social and Cognitive Networks Academic Research Center,
Rensselaer Polytechnic Institute, 110 8$^{th}$ Street, Troy, NY 12180--3590, USA}
\affiliation{Department of Computer Science, Rensselaer Polytechnic Institute,
110 8$^{th}$ Street, Troy, NY 12180--3590, USA}
\affiliation{Department of Physics, Applied Physics, and Astronomy,
Rensselaer Polytechnic Institute, 110 8$^{th}$ Street, Troy, NY 12180--3590, USA}

\author{J. Xie}
\affiliation{Social and Cognitive Networks Academic Research Center,
Rensselaer Polytechnic Institute, 110 8$^{th}$ Street, Troy, NY 12180--3590, USA}
\affiliation{Department of Computer Science,
Rensselaer Polytechnic Institute, 110 8$^{th}$ Street, Troy, NY 12180--3590, USA}

\author{B.K. Szymanski}
\affiliation{Social and Cognitive Networks Academic Research Center,
Rensselaer Polytechnic Institute, 110 8$^{th}$ Street, Troy, NY 12180--3590, USA}
\affiliation{Department of Computer Science,
Rensselaer Polytechnic Institute, 110 8$^{th}$ Street, Troy, NY 12180--3590, USA}

\author{G. Korniss}
\affiliation{Social and Cognitive Networks Academic Research Center,
Rensselaer Polytechnic Institute, 110 8$^{th}$ Street, Troy, NY 12180--3590, USA}
\affiliation{Department of Physics, Applied Physics, and Astronomy,
Rensselaer Polytechnic Institute, 110 8$^{th}$ Street, Troy, NY 12180--3590, USA}


\begin{abstract}
We investigate consensus formation and the asymptotic consensus
times in stylized individual- or agent-based models, in which global
agreement is achieved through pairwise negotiations with or without
a bias.
Considering a class of individual-based models on finite complete
graphs, we introduce a coarse-graining approach (lumping microscopic
variables into macrostates) to analyze the ordering dynamics in an
associated random-walk framework.
Within this framework, yielding a linear system, we derive general
equations for the expected consensus time and the expected time
spent in each macro-state. Further, we present the asymptotic
solutions of the 2-word naming game, and separately discuss its
behavior under the influence of an external field and with the
introduction of committed agents.
\end{abstract}


\maketitle

{\bf Individual- or agent-based models provide invaluable tools to
investigate the collective behavior and response of complex social
systems. These systems typically consists of a large number of
individuals interacting through a random and sparse network
topology. Despite this non-trivial network topology, it has been
demonstrated in many recent examples that unlike in low-dimensional
spatially-embedded systems with short-range connections, the
collective dynamics in sparse random graphs (with no community
structure) exhibit scaling properties very similar to those observed
on the complete graph. Therefore, studying fundamental agreement
processes on the complete graph can yield insights for the ordering
process in more realistic sparse random networks. In this paper we
consider two simple individual-based models and develop a
mathematical framework which yields asymptotically exact consensus
times for large but finite complete graphs of size $N$. In
particular, after demonstrating the feasibility of this framework on
known examples, we apply it to study two distinct stylized
approaches in social influencing: (i) influencing individuals by a
global external field (mimicking mass media effects) and (ii)
introducing committed individuals with a fixed designated opinion
(who can influence others but themselves are immune to influence).
In the former case, we find the external field dominates the
consensus in the large network-size limit, while in the latter case
we find the existence of a tipping point, associated with the
disappearance of the metastable state in the opinion space. The
results further our understanding of timescales
associated with reaching consensus in social networks.}

\section{Introduction}
Research on non-equilibrium models for social and opinion dynamics
has attracted considerable attention \cite{Castelano_RMP2009,Galam_IJMPC2008,Pan2009}. In
this paper we present a method to calculate consensus (or ordering)
times for a large class of individual-based models with discrete
state-variables on the complete graph of size $N$. The mode of
reaching consensus depends on the initial configuration and whether
or not the individuals are under the influence of a bias. For
example, starting from a configuration of evenly mixed opinions with
no bias, these systems can exhibit ``coarsening" in which case the
consensus time typically diverges with the system size in a
power-law fashion
\cite{Pan2009,Krapivksy_PRA1992,Krapivsky_PRE1996,Ben-Naim1996,Dornic_PRL2001,Baronchelli2006-2,Lu_PRE2008}.
On the other hand, initializing the system in one particular opinion
and exerting some weak bias favoring another opinion create systems that
undergo an ``escape" from the meta-stable state \cite{Lucomm}. For
networks of finite size, it is important to characterize the consensus time, i.e. the time
until the system fully orders.

Direct simulations of the above behaviors can be time-consuming for
large systems and essentially impossible even for moderately-sized
systems when the system initially is in a meta-stable configuration,
as the escape time can increase exponentially with the system size.
The method presented here provides a way to obtain the asymptotic
behavior of consensus times, including the cases associated with
extremely slow meta-stable escapes.
While the method and the results presented here are applicable to
the complete graph, consensus times often exhibit the same
asymptotic scaling with $N$ in large homogeneous sparse random
networks \cite{Castelano_PRE2005,DallAsta_PRE2006,Lu_PRE2008}, hence
one can gain some insight how ordering and consensus evolves in
realistic social networks.

The Naming Game (NG) first emerged in linguistic modeling of the
transition from microscopic local interactions to global consensus
in the absence of a global coordinator
\cite{Pan2009,Baronchelli2006-2,Steels_1995}. This phenomena of total
synchrony or global consensus has general interest and applications
such as sensor networks, some aspects of neuro-science models for
brain functions and social dynamics. Earlier research in this domain
mainly discussed two aspects of this model. One focuses on the
relationship between the network topology and the dynamics
\cite{Baronchelli2006,DallAsta2006,DallAsta_PRE2006}, the other, in contrast,
ignores all the effects of network topology and studies the
intrinsic property of the dynamics on the complete graph
\cite{Baronchelli2006-2,Baronchelli_IJMPC2008}. The latter for large
network size $N$ (also referred to as mean-field or homogeneous
mixing) is what we consider in this paper. We assume here an
initial configuration where every agent has a non-empty name list of
words. The dynamics on the complete graph is driven by the mean
field (in this context referring to the fraction of nodes in each
state) rather than the detailed network states (including the states
of all the nodes). It naturally leads to a coarse-graining approach
by lumping microscopic variables into macro-states. Our
paper studies the impact of a global external field
(mimicking mass media effects) and of presence of committed
agents (who will firmly stick to and convey a designated opinion) on
the consensus process.

A point worth mentioning is that the Naming Game constitutes an opinion
dynamics model in which in which a node can possess multiple
opinions simultaneously. Such models with intermediate states have only
recently begun receiving attention \cite{Dallasta_2008,Vazquez_2010} in statistical physics literature, and we
believe that this study is an important contribution to this body of literature.


The paper is organized as follows. In Secs.~II and III, we introduce
the coarse-graining approach to map models for opinion dynamics on
the complete graph to an associated random walk problem, and test
our framework through known results for the spontaneous agreement
process (without any influencing)
\cite{Slanina_EPJB2003,Sood_PRE2008,Castelano_PRE2005,Baronchelli_IJMPC2008,Castello_EPJB2009}.
We derive the equations in a linear-system form for the expected
total consensus time and the expected time spent in each macro-state
which can be solved in a closed form for the voter model
\cite{Liggett_1985,Castelano_RMP2009,Krapivksy_PRA1992} (Sec.~II).
Then, in  Sec.~III, we employ the same method for a variation of the
Naming Game (NG) with two words \cite{Castello_EPJB2009}. Finally,
and most importantly, we investigate models for social influencing:
we present new asymptotic results and discuss the behavior of the
2-word NG under the influence of an external field (Sec.~IV) and in the
presence of committed agents (Sec.~V). A brief summary is given in
Sec.~VI.

\section{Consensus time in the voter model on the complete graph}

First, we consider a well studied prototypical model for opinion
formation, the voter model
\cite{Liggett_1985,Castelano_RMP2009,Krapivksy_PRA1992}. In this model, the
evolution of a suitably defined global variable can be easily mapped
onto a random-walk problem. Further, the solution of this model is
known in all dimensions, including the complete graph, hence our
method can be tested.

Given a network of $N$ nodes, with each node in a state chosen from the set
of possible opinions $X$, the voter model is defined by the following update rule:


\emph{A pair of nodes, a ``speaker" and a ``listener", are chosen at random.
The listener then changes its state to that of the speaker.}

If the set of nodes in the network is denoted by $S$, ($\left |{S}\right |$),
then the above rule defines a Markov chain in an $N$ dimensional space  $X^S$.
Under mean field assumption, which is justifiable when
dealing with complete graphs, one coarse-graining
approach is to take all network states in $X^S$ corresponding to the same mean field $\vec{n}=(n_1,n_2,...,n_M)$ as
a macrostate, where $n_i$ denotes the number of nodes in state $i\in
X$ and $M = \left |{S}\right |$. Therefore, the coarse-grained
random process is valued in a $M-1$ hyper-plane (since $\sum_{i=1}^M n_i=N$)
in M dimension space. When $M=2$ (2-state voter model), taking
$X=\{A,B\}$, the coarse-grained process is on a segment (since $n_A+n_B=N
(n_A, n_B>=0)$), and all macrostates can be represented by a single
discrete variable $n_A=0,1,...,N$.

In each time step $\D n_A$, the change of $n_A$, is a random
variable depends only on current macrostate $n_A$. Its possible
values and the corresponding events and probabilities are listed Table~\ref{table:1}.
\begin{table}
\caption{Update events for the voter model and the associated random walk transition probabilities}
\begin{center}
\begin{tabular}{c|c|c|c|l}\hline
speaker& listener & event & $\D n_A (n_A)$ & probability\\ \hline
A & B &$B\go A$ & 1 & $(1-{n_A\over N}){n_A \over N-1}$\\
B & A &$A\go B$ & -1 & ${n_A\over N}(1-{n_A-1 \over N-1})$\\
A & A &unchanged & 0 & ${n_A\over N}{n_A \over N-1}$\\
B & B &unchanged & 0 & $(1-{n_A\over N})(1-{n_A-1 \over N-1})$\\\hline
\end{tabular}
\end{center}
\label{table:1}
\end{table}

Hence, the coarse-grained process of the voter model can be mapped to a random walk in 1-d,
\begin{equation}
n_A(T)=n_A(0)+\sum_{t=0}^{T-1} \D n_A(n_A(t)) \;,
\end{equation}
where $n_A=0,N$ are two   absorbing states.


\subsection{First-step analysis of consensus times in the voter model}
The expected time before absorption which is the quantity of the
most interest can be evaluated by {\it first-step analysis}
\cite{Bremaud}.
The idea is based on a straight forward statement: The absorption
time can be decomposed into two parts, the time steps before  and
after leaving the current macrostate. The former one is called the
residence time $t_r(n_A)$ of the given macrostate $n_A$. We denote
the expected time before absorption and the expected residence time
starting from a specific macrostate $n_A$ on complete graphs with
$N$ nodes as $\tau(n_A|N)$ and $t(n_A|N)$, respectively, or simply
$\tau(n_A)$ and $t(n_A)$ when it is not ambiguous. Then taking
expectation of all random variables mentioned in the statement
above, we get the following equations:
\begin{eqnarray}
\tau(n_A) & = & E[\mbox{time before leaving $n_A$}]+ E[\mbox{time before absorption and after leaving $n_A$}] \nonumber \\
              & = & E[t_r(n_A)] + E[E[\mbox{time before absorption and after leaving $n_A$}|\mbox{entering new macrostate $n_A'$}]] \nonumber \\
              & = & t(n_A) + \frac{\sum_{i=-1,1} P(\D n_A(n_A)=i)\tau(n_A+i))}{ 1-P(\D n_A(n_A)=0)}
              \; = \; t(n_A) + \frac{1}{2}(\tau(n_A+1)+\tau(n_A-1)) \;,
\label{timetoabsorb}
\end{eqnarray}
where $t(n_A)$ is given by following argument:
\begin{eqnarray} t(n_A) &=&\Sigma_{k=1}^\infty P(t_r(n_A)=k) k  \; = \; \Sigma_{k=1}^\infty P(t_r(n_A)\geq k) \nonumber \\
            &=&\Sigma_{k=1}^\infty P(\D n_A(n_A)=0)^{k-1}       \; = \; {1\over 1-P(\D n_A(n_A)=0)}
            \; = \; {1\over 2 {n_A\over N} (1-{n_A\over N-1})} \;.
\end{eqnarray}
Defining $\vec{\tau}=(\tau(1),...,\tau(N-1))^T$,
$\vec{t}=(t(1),...,t(N-1))^T$ and using boundary conditions
$\tau(0)=\tau(N)=0$, we rewrite Eq.~(\ref{timetoabsorb}) as a linear
system:
\begin{equation}
\vec{\tau}=Q\vec{\tau}+\vec{t} \;,
\end{equation}
where
\begin{equation}
Q=\left[\begin{array}{ccccc}
                           0 & \half &  &  &            \\
                           \half & \cdot & \cdot &  &   \\
                            & \cdot & \cdot & \cdot &   \\
                            &  & \cdot & \cdot & \half  \\
                           & & & \half & 0  \end{array}
\right]_{(N-1)\times (N-1)} \;.
\end{equation}
$\tau(n_A|N)$ can be solved exactly from this equation in terms of $n_A$ and $N$. Define $\d_N^i$ as a N-entries column vector in which only the i\emph{th} entry is non-zero and has value 1. We have:
\begin{equation}
     \tau(i|N)     = (\d_{N-1}^{i})^T \vec{\tau}
                \; = \; (\d_{N-1}^{i})^T (I-Q)^{-1} \vec{t}
                \; = \; \vec{u}\cdot \vec{t} \;,
\end{equation}
where $\vec{u}=(u_1,u_2,...,u_j,...,u_{N-1})$ is the solution of equation $(I-Q)^T\vec{u}=(\d_{N-1}^{i})$. Intuitively, $u_j$ can be understood as the average number of visits (assuming entering and leaving a macrostate as one visit) of macrostate $n_A=j$ before absorption and starting from macrostate $n_A=i$. Moreover $T_j=u_j t_j$
is actually the average number of time steps (including self-repeating steps) spent on the macrostate $n_A=j$ before absorption. It is quite easy to show that $\tau(n|N)$ is monotonic with respect to $N$. Therefore, in order to obtain the large $N$ behavior of $\tau(n_A|N)$ we focus on the case when $N$ is even and always assume $n_A(0)=N/2$. In such cases,
$\vec{u}=(1,2,3,...,N/2,...3,2,1)^T$ and

\begin{eqnarray}
\tau(N/2|N)&=&\vec{v_{N/2}} \cdot \vec{t} \;=\; 2\Sigma_{k=1}^{N/2-1} k {1\over 2 (k/N) (1-k/(N-1))}+ {N\over 2} 2 \nonumber \\
                &=& N(N-1)\Sigma_{k=1}^{N/2-1}{1\over N-1-k}+N \;\approx\; N(N-1) \int_{N/2}^{N-2} {dx\over x} + N \nonumber \\
                &\approx& \ln(2) N(N-1) + N \;.
\end{eqnarray}
As is the convention for agent-based models in statistical physics,
unit time is assumed to consist of $N$ update events. Thus,
following this convention, the normalized consensus time is
$\tau(N/2|N)/N$ is $\ln(2)N$ to the leading order. This agrees with
the scaling behavior obtained previously for the characteristic
(relaxation) time of the voter model through a Fokker-Planck
\cite{Slanina_EPJB2003,Sood_PRE2008} approach and through
simulations \cite{Castelano_PRE2005}. .

\begin{center}
\begin{figure}[!htbp]
  \includegraphics[width=0.8\textwidth]{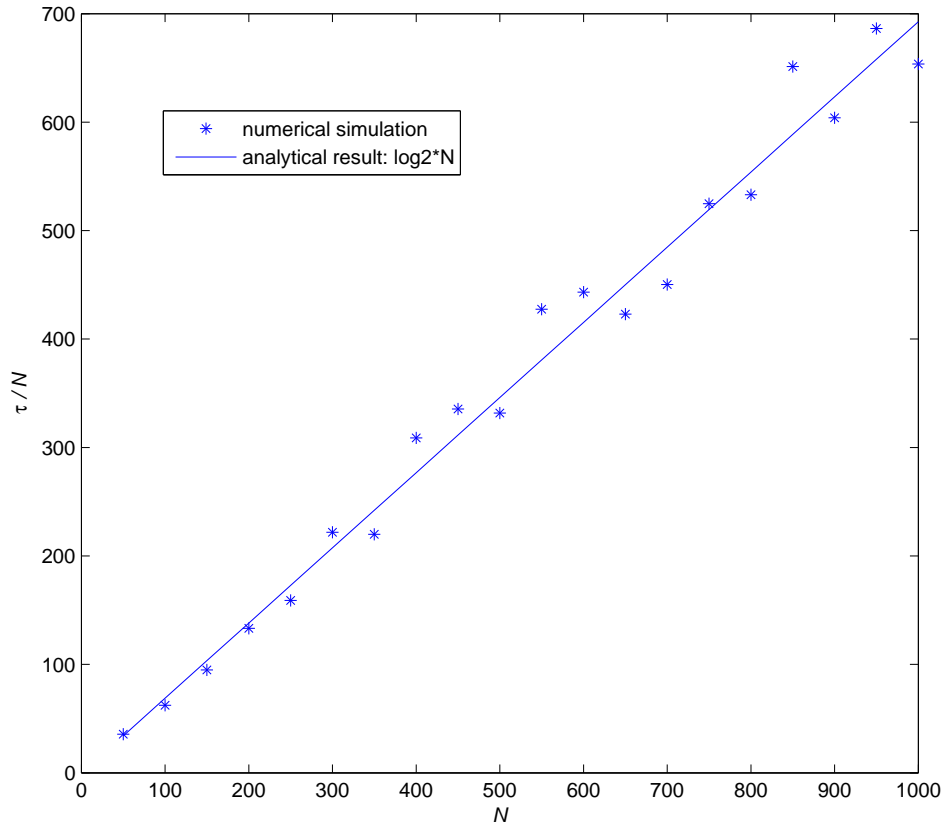}\\
  \caption{Consensus time for voter model on complete graph. The vertical axis represent the consensus time (normalized by $N$). The horizontal axis is the number of nodes in complete graph. Each star
  point is an average of 10 runs of numerical simulations of voter model and the solid straight line consists of the solutions of the linear
  equation for each $N$ value.}
  \label{figure:1}
\end{figure}
\end{center}

Figure \ref{figure:1} shows the comparison of the consensus time from
numerical simulations (averaged over 50 runs for each $N$) with the analytical results.

\section{The 2-word Naming Game}
The Naming Game (NG) \cite{Baronchelli2006,Baronchelli_IJMPC2008,Castello_EPJB2009} is
somewhat more complicated than the voter model because for the given set
of all possible opinions (words in the original NG) $X$ $(|X|=M)$, the state
of each node is a member of the power set $2^X$ - the set of all subsets of $X$  - rather than $X$
itself. Moreover, the update rule is replaced by:

\emph{A pair of nodes, a ``speaker" and a ``listener", are chosen at random.
The speaker then randomly selects one word from her list and communicates it to the listener.
If the listener already has this word (termed ``successful communication"), she deletes all other words in her list
(i.e., collapses her list to this most recently communicated word);
If the listener does not have the word communicated by the speaker, she adds it to her list (hence, individuals can
carry more than one word at a time).}

The slight difference between the update rule defined above and that of the original NG \cite{Baronchelli2006}
is that here, upon ``successful" communication, {\em only the listener changes its state}.
This restriction eliminates steps of size 2 in the associated RW model, making it easier to apply the method developed in Sec.~II.A.
Furthermore, we will consider the version of the NG with only two words \cite{Castello_EPJB2009}.

\begin{center}
\begin{figure}[htbp]
  \includegraphics[width=0.8\textwidth]{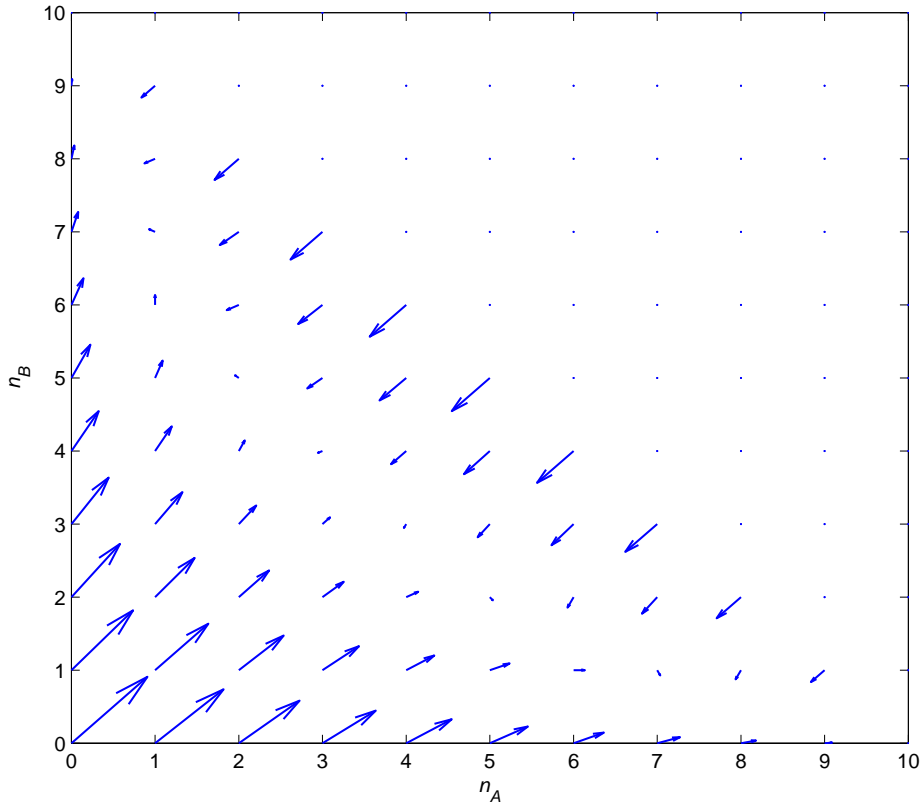}\\
  \caption{Vector field $E[\D \vec{n}(\vec{n})]$ of the random walk coarse-grained from 2-word naming game. Each vector is the expected drift of  the random walk at macrostates $\vec{n}$. The network is 10 nodes complete graph and the domain of random walk is the lower triangle of the square lattice}
  \label{figure:2}
\end{figure}
\end{center}

The coarse-graining approach
mentioned in Section II merges all network states labeled by the same
vector $\vec{n}=(n_i)_{2^M-1}, i\in 2^X\backslash \{\emptyset\}$ into
a macrostate. Here $n_i$ is the number of nodes in state $i$. The
coarse-grained random process takes values in the $2^M-2$
hyper-plane: $\sum_{i\in2^X\backslash \{\emptyset\}}n_i=N$, thus we
can map the coarse-grained process into a $2^M-2$ dimension
random walk. In the case of 2-word NG $M=2$, so assuming
$X=\{A,B\}$,  $\vec{n}=(n_{\{A\}},n_{\{B\}},n_{\{A,B\}})$ or
$(n_A,n_B,n_{AB})$ for short, where $n_A$, $n_B$, and $n_{AB}$ are the number
of individuals carrying word A, word B, or both, respectively.
Since $n_{AB}=N-n_A-n_B$, we dump one
redundant dimension and take the 2-d vector $\vec{n}=(n_A, n_B)$ to
represent the macrostate. In each time step, the change of
macrostate $\D \vec{n}$ have five possible values. For all these possible
values of $\D \vec{n}$, the corresponding events and probabilities
are listed in Table~\ref{table:2}.\\
\begin{table}
\caption{Update events for the 2-word naming game and the associated random walk transition probabilities}
\begin{center}
\begin{tabular}{c|c|c|c|l}\hline
speaker&listener & event &$\D \vec{n}(n_A,n_B)$  & probabilty \\ \hline
B or AB &A   &$A\go AB$ &(-1,0) & $P(A-)=n_A (N-n_A+n_B)/2N^2$\\
A or AB &AB  &$AB\go A$ &(1,0)  & $P(A+)=(N-n_A-n_B)(N+n_A-n_B)/ 2N^2$\\
A or AB &B   &$B\go AB$ &(0,-1) & $P(B-)=n_B (N+n_A-n_B)/ 2N^2$\\
B or AB &AB  &$AB\go B$ &(0,1)  & $P(B+)=(N-n_A-n_B)(N-n_A+n_B)/ 2N^2$\\
A, B or AB &A or B &unchanged &(0,0) & $P_0=(n_A+n_B)/2N+(n_A-n_B)^2/2N^2$\\\hline
\end{tabular}
\end{center}
\label{table:2}
\end{table}
In Table~\ref{table:2}, the event $A\go AB$, for example, means the listener node changes its state from A to AB.
Analogous to the procedure followed for the voter model, we map the coarse-grained 2-word NG to a 2-d random walk,
\begin{equation}
\vec{n}(T)=\vec{n}(0)+\sum_{t=0}^{T-1} \D \vec{n}(\vec{n}(t)) \;.
\end{equation}
Here the domain $D$ of $\vec{n}$ is the set containing all integer grid points bounded by the lines $n_A=0$, $n_B=0$ and $n_A+n_B=N$
while $\vec{n}=(0,N),(N,0)$ are two absorbing states.
The expected drifts, $E[\D
\vec{n}(\vec{n})]=(P(A+)-P(A-),P(B+)-P(B-))$ are plotted in
Fig.~\ref{figure:2} for $N=10$. As shown in Fig.~\ref{figure:2}, on
average, $\vec{n}(t)$ will quickly go to a stable trajectory and
then slowly converge to one of the consensus states. In other words,
in contrast to an unbiased random walk as in the voter model, the
2-word NG is "attracted" to the consensus state after a spontaneous
symmetry breaking. So it is reasonable to expect that the 2-word NG
will achieve consensus much faster than the voter model,
both starting from the $n_A=n_B=N/2$ initial configuration on the complete graph.

\subsection{First-step analysis of consensus times in the 2-word Naming Game}
We now repeat the first-step analysis (developed in Sec.~II), noting that the method is essentially independent of the number of dimensions.
Assume $\tau(\vec{n}|N)$ and $t(\vec(n)|N)$ are expected numbers of time steps before absorption and before leaving the current state starting
at the macrostate $\vec{n}=(n_A,n_B)$, then we have:
\begin{eqnarray}
\tau(\vec{n})&=& t(\vec{n})+{\sum_{i\in \{(1,0),(-1,0),(0,1),(0,-1)\}} P(\D \vec{n}(\vec{n})=i)\ \tau(\vec{n}+i)\over 1-P(\D \vec{n}(\vec{n})=(0,0))} \nonumber \\
                  &=& t(\vec{n})+{P(A+)\tau(n_A+1,n_B)+P(A-)\tau(n_A-1,n_B)+P(B+)\tau(n_A,n_B+1)+P(B-)\tau(n_A,n_B-1)\over 1-P(0)}
\label{timetoabsorb2}
\end{eqnarray}
and
\begin{eqnarray}
t(\vec{n}) ={1\over 1-P(\D \vec{n}(\vec{n})=(0,0))} ={1\over 1-(n_A+n_B)/2N-(n_A-n_B)^2/2N^2}
\end{eqnarray}
Ordering all the macrostates $\vec{n}$ (except the two absorbing states)
in a vector, and arranging $\tau(\vec{n})$,$t(\vec{n})$ in the same
order to get $\vec{\tau}$ and $\vec{t}$ whose dimensions are
$(N+2)(N+1)/2-2$, we write the Eq.~(\ref{timetoabsorb2}) in the same linear
system form $\vec{\tau}=Q\vec{\tau}+\vec{t}$. Furthermore, taking
$\d_N^{\vec{n}}$ as a column vector where the only nonzero entry is
1 corresponding to $\vec{n}$, we solve
expected number of visits to each macrostate through the equation
$(I-Q)^T u=\d_N^{\vec{n}}$. The expected number of time steps spent
on each macrostate $T(n_A,n_B)$ is obtained by multiplying corresponding elements of $u$ and
$\vec{t}$. Although in this case the matrix $Q$ is not easy to write
generally, it has a good property that sum of each row is
$\leq 1$ and for some rows (those corresponding to
$\vec{n}=(N-1,0)$ and $(0,N-1)$) the inequality is strict. Consequently
the moduli of all the eigenvalues of $Q$ is strictly less than 1 and
$(I-Q)$ is invertible, therefore the existence and uniqueness of the
solutions are guaranteed.

\begin{center}
\begin{figure}[htbp]
  \includegraphics[width=0.8\textwidth]{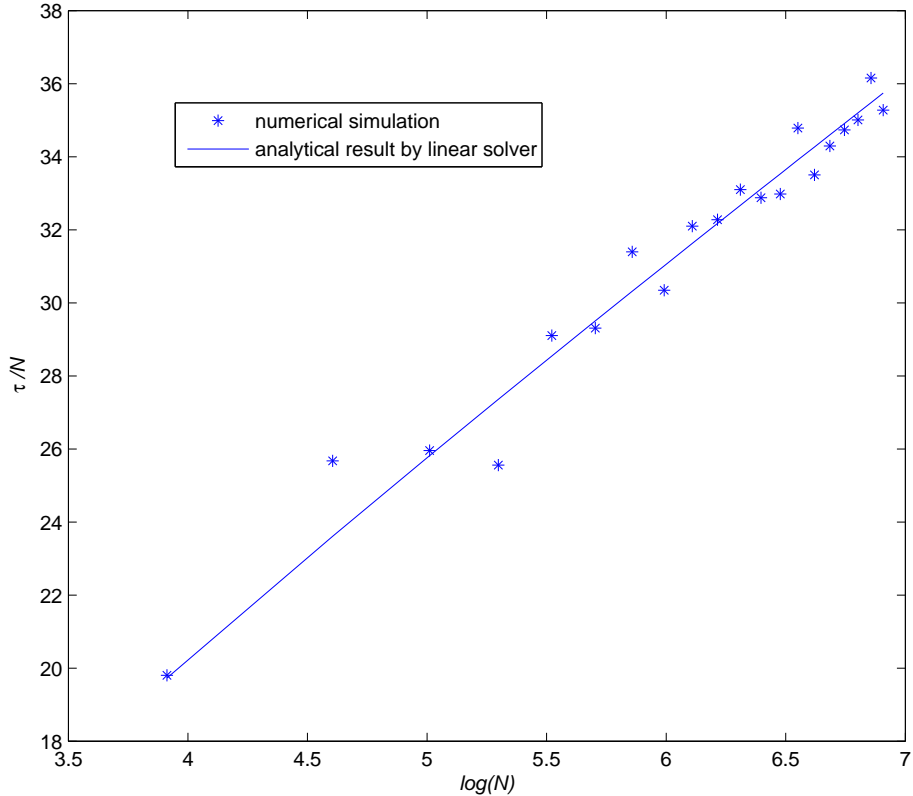}\\
  \caption{Consensus time (normalized by $N$) as a function of network size $N$ for 2-word naming game on complete graph. Each star
  point is an average of 10 runs of numerical simulations of 2-word naming game and the solid straight line consists of the solutions of the linear
  equation for each $N$.}
  \label{figure:3}
\end{figure}
\end{center}

\begin{center}
\begin{figure}[htbp]
  \includegraphics[width=0.7\textwidth]{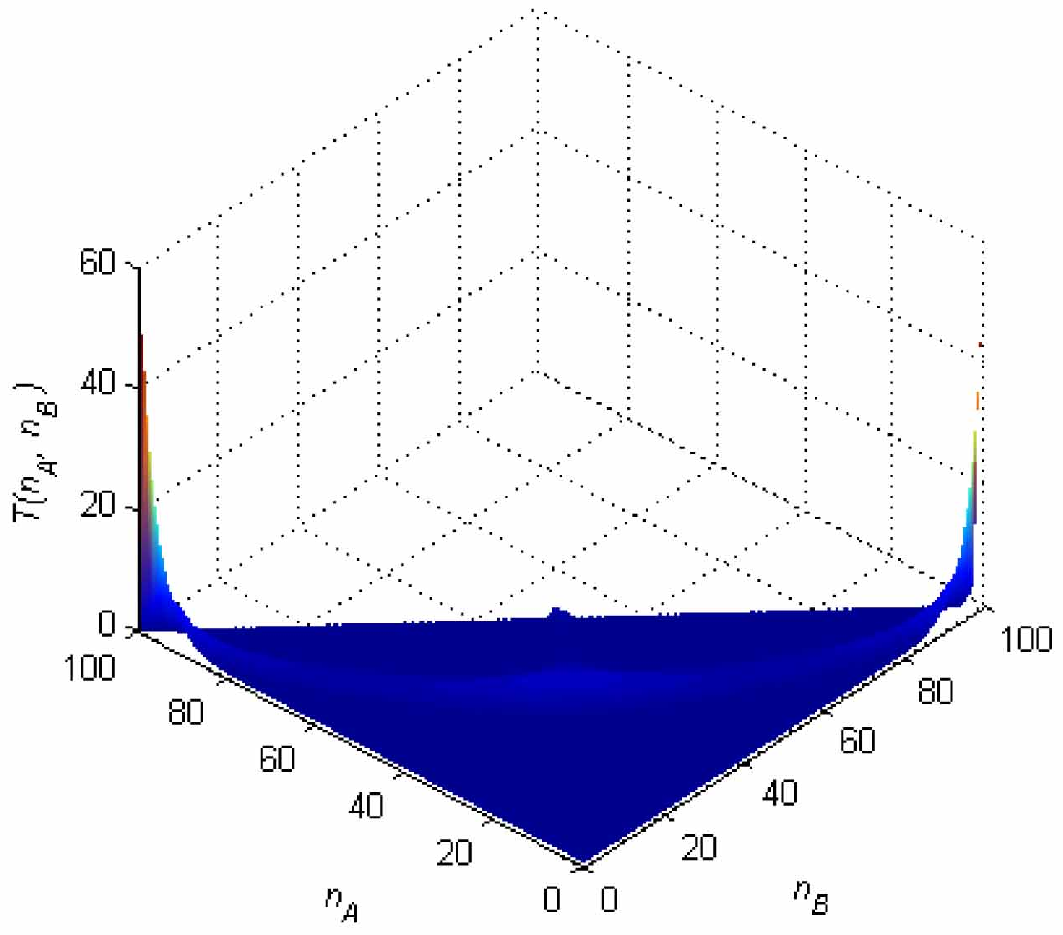}\\
  \caption{The expected time spent on each macrostate before consensus in the 2-word NG on a complete graph with $N=100$ nodes.
  The vertical axis $T(n_A,n_B)$ is the expected time that the random walk spends in macrostate $(n_A, n_B)$ before
  consensus, starting from the $(n_A(0),n_B(0))=(50,50)$ initial macrostate.}
  \label{figure:4}
\end{figure}
\end{center}

We study the case of even $N$ and unbiased initial state
$\vec{n}(0)=(N/2,N/2)$, and compare the normalized consensus time
$\tau(N/2,N/2|N)$ obtained from numerical simulations against the solution obtained from the
linear system. Figure \ref{figure:3} shows that the normalized
consensus time has an order $O(\ln{N})$ which is smaller than the
order $O(N)$ for the voter model. Note that the $O(\ln{N})$
consensus time of the original 2-word NG with the above initial
configuration has been previously found using simulations and a rate
equation approach \cite{Baronchelli_IJMPC2008,Castello_EPJB2009}.
Figure \ref{figure:4}, depicting the expected number of time steps
spent on each macrostate $T(n_A,n_B)$ when $N=100$ and
$\vec{n}(0)=(50,50)$ shows that the time before absorption is mainly
spent on the macrostates near the consensus states.


\section{The 2-word naming game with external influence}
The previous section focused on the time taken by the system to
spontaneously reach consensus. A natural question that arises is how
consensus can be sped up through an external influencing force such
as mass media \cite{Mazitello_IJMP2007,Candia_JSTAT2008}, In this
section we study this case: the 2-word NG subject to an external
field of magnitude $f$ for which the update rule is defined as
follows: in each time step, if the listener is in a mixed state i.e.
AB, it will with probability $f$ spontaneously change into state A
and with probability $1-f$ follow the original NG rule (Sec.~III).
All the differences between this case and the spontaneous case lie
in the $\D \vec{n}(\vec{n})$ which are listed in
Table~\ref{table:3}.
\begin{table}
\caption{Update events for the 2-word naming game with central influence and the associated random walk transition probabilities}
\begin{center}
\begin{tabular}{c|c|c|c|l}\hline
speaker&listener & event &$\D \vec{n}(n_A,n_B)$  & probabilty \\ \hline
B or AB &A   &$A\go AB$ &(-1,0) & $P(A-)=n_A (N-n_A+n_B)/2N^2$\\
A, AB or f &AB  &$AB\go A$ &(1,0)  & $P(A+)=(1-f)(N-n_A-n_B)(N+n_A-n_B)/ 2N^2+f(N-n_A-n_B)/N$\\
A or AB &B   &$B\go AB$ &(0,-1) & $P(B-)=n_B (N+n_A-n_B)/ 2N^2$\\
B or AB &AB  &$AB\go B$ &(0,1)  & $P(B+)=(1-f)(N-n_A-n_B)(N-n_A+n_B)/ 2N^2$\\
A, B or AB &A or B &unchanged &(0,0) & $P_0=(n_A+n_B)/2N+(n_A-n_B)^2/2N^2$\\\hline
\end{tabular}
\end{center}
\label{table:3}
\end{table}

\begin{center}
\begin{figure}[htbp]
  \includegraphics[width=0.8\textwidth]{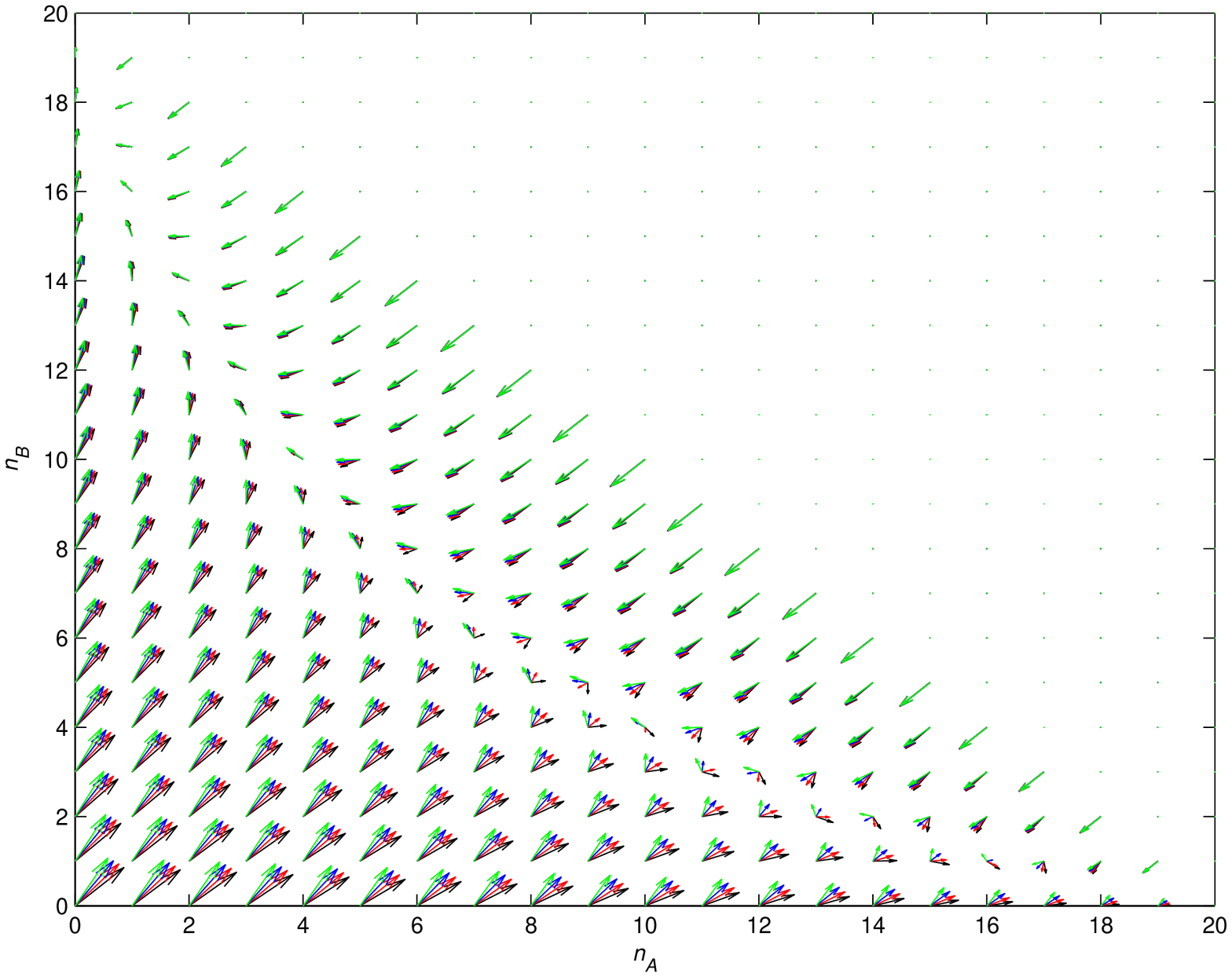}\\
  \caption{Several vector fields are drawn in this figure with different colors,
   each color show the drift of coarse-grained random walk $E[\D \vec{n}(\vec{n}) ]$ on a complete graph with $N=20$ nodes
   at different central influence levels $f$. The length of each vector has been rescaled to its square root to avoid cluttering the whole graph.}
  \label{figure:5}
\end{figure}
\end{center}

In Fig.~\ref{figure:5} we show how the vector field $E[\D n]$ which
is intuitively the ``drift" part of the course-graining random walk
changes at different influence levels $f$'s.

\begin{center}
\begin{figure}[htbp]
  \includegraphics[width=0.7\textwidth]{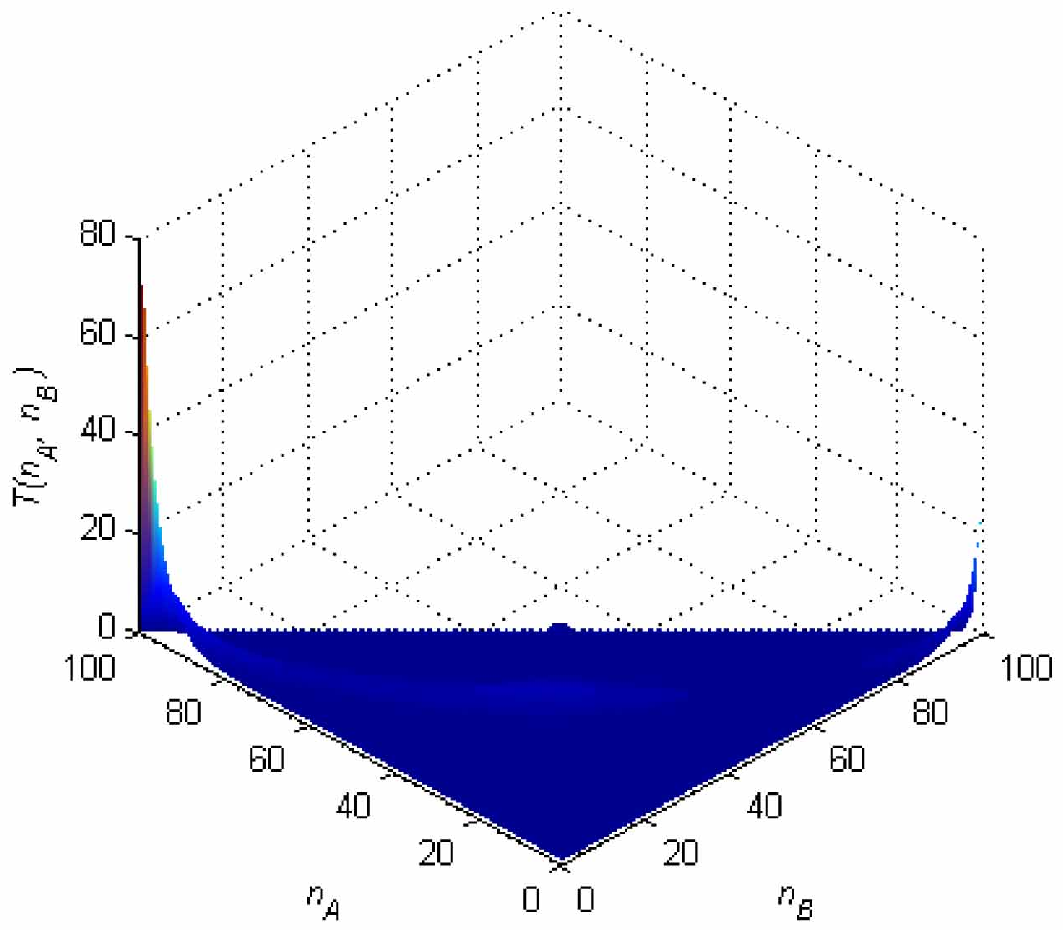}\\
  \caption{Expected time spent on each macrostate before consensus $T(n_A,n_B)$ in the 2-word NG on complete graph with $N=100$
  with central influence $f=0.05$, starting from a unbiased initial macrostate $(n_A(0),n_B(0))=(50,50)$.}
  \label{figure:6}
\end{figure}
\end{center}

Following the first-step analysis, we can solve for the expected consensus
time $\tau$ and the expected number of time steps spent at each
macrostate $T(n_A,n_B)$ starting from any given macrostate. A
specific solution of $T(n_A,n_B)$ on a complete graph with $N=100$
starting from macrostate $(50,50)$ is shown in Fig.~\ref{figure:6}.
As shown, there are two peaks around the two consensus states, just
as in the spontaneous case, although the peak near the consensus
state which the external influence prefers (all A state) is much
higher than the other one, even at a low influence level $f=0.05$.

\subsection{First-step analysis of probability of consensus}
To better understand the effect of external influence, we apply the first-step analysis on the probability of reaching a specific consensus state. Defining $P_A(\vec{n})$ as the probability of going to an all-$A$ consensus starting from the macrostate $\vec{n}=(n_A,n_B)$, $P_A(\vec{n})$ follows:
\begin{eqnarray}
P_A(\vec{n}(t))= E[\mbox{$P_A(\vec{n}(t+1))$}]={\sum_{i\in \{(1,0),(-1,0),(0,1),(0,-1),(0,0)\}} P(\D \vec{n}(\vec{n})=i)\ P_A(\vec{n}+i)}
\end{eqnarray}
and satisfies the boundary conditions $P_A(N,0)=1$ and $P_A(0,N)=0$. Ordering $P_A(\vec{n})$ of all macrostates (including the two consensus states) in a vector $\vec{P_A}$, we rewrite the equations as $\vec{P_A}=Q_0 \vec{P_A}$. $Q_0$ is a square matrix of order $(N+2)(N+1)/2$ and all its elements are given in Table III.

\begin{center}
\begin{figure}[htbp]
  \includegraphics[width=0.8\textwidth]{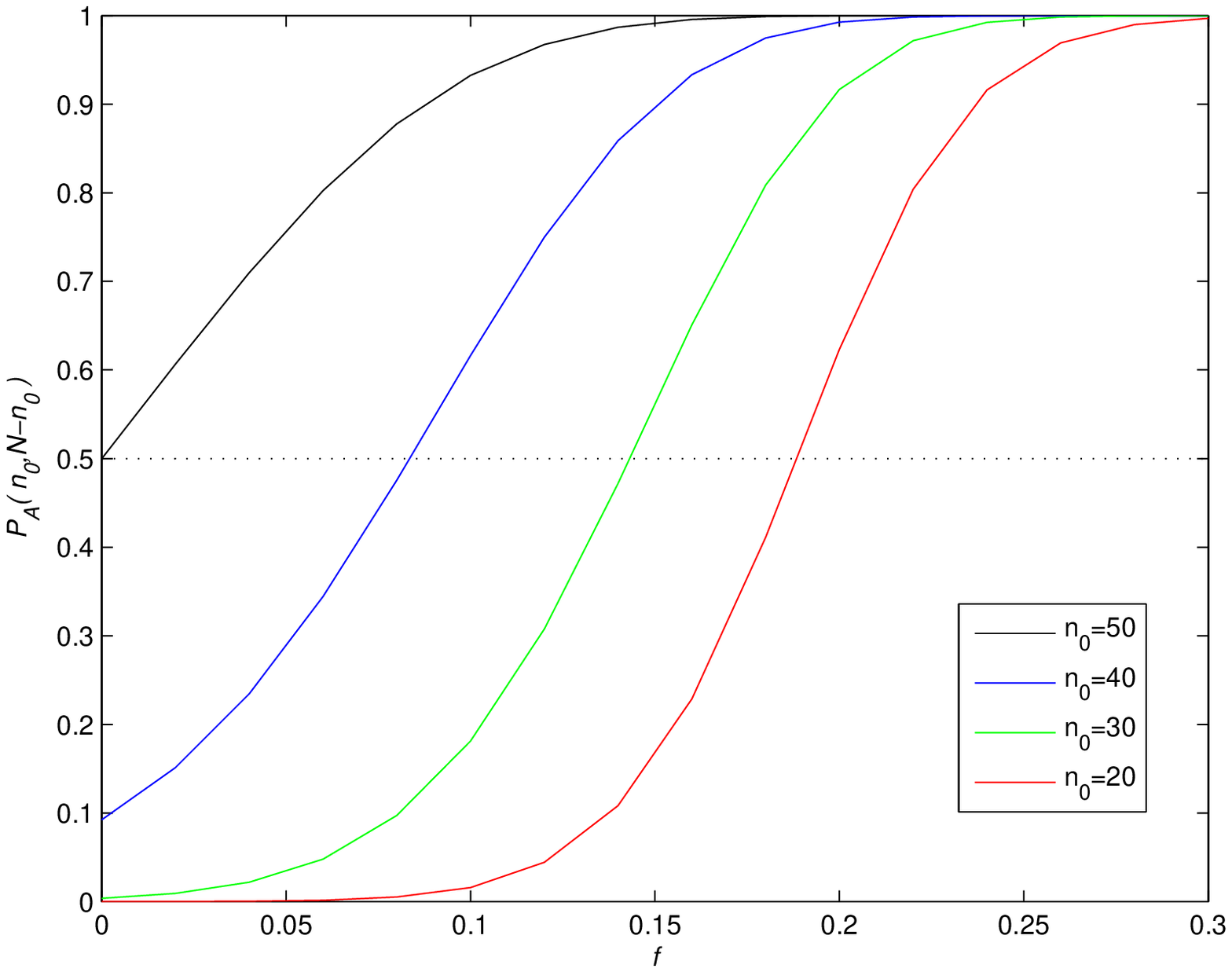}\\
  \caption{Probability of all A consensus $P_A$ with different external influence level $f$'s starting from macrostate $(n_A,n_B)=(n_0,N-n_0)$ on 100-node complete graph.}
  \label{figure:7}
\end{figure}
\end{center}

In Fig.~\ref{figure:7}, we consider the NG on 100-node complete
graph starting from the macrostate $(n_0, N-n_0)$. For $n_0=50$ and
central influence $f=0$ (the left end of the black curve), it has
equal probability of going to all A and all B consensus. When
$n_0<50$, it is more probable to go to all B consensus without
central influence. However, with a biased central influence $f$, one
can always convert the preference of the process to the opposite
side - all A consensus.

\begin{center}
\begin{figure}[htbp]
  \includegraphics[width=0.8\textwidth]{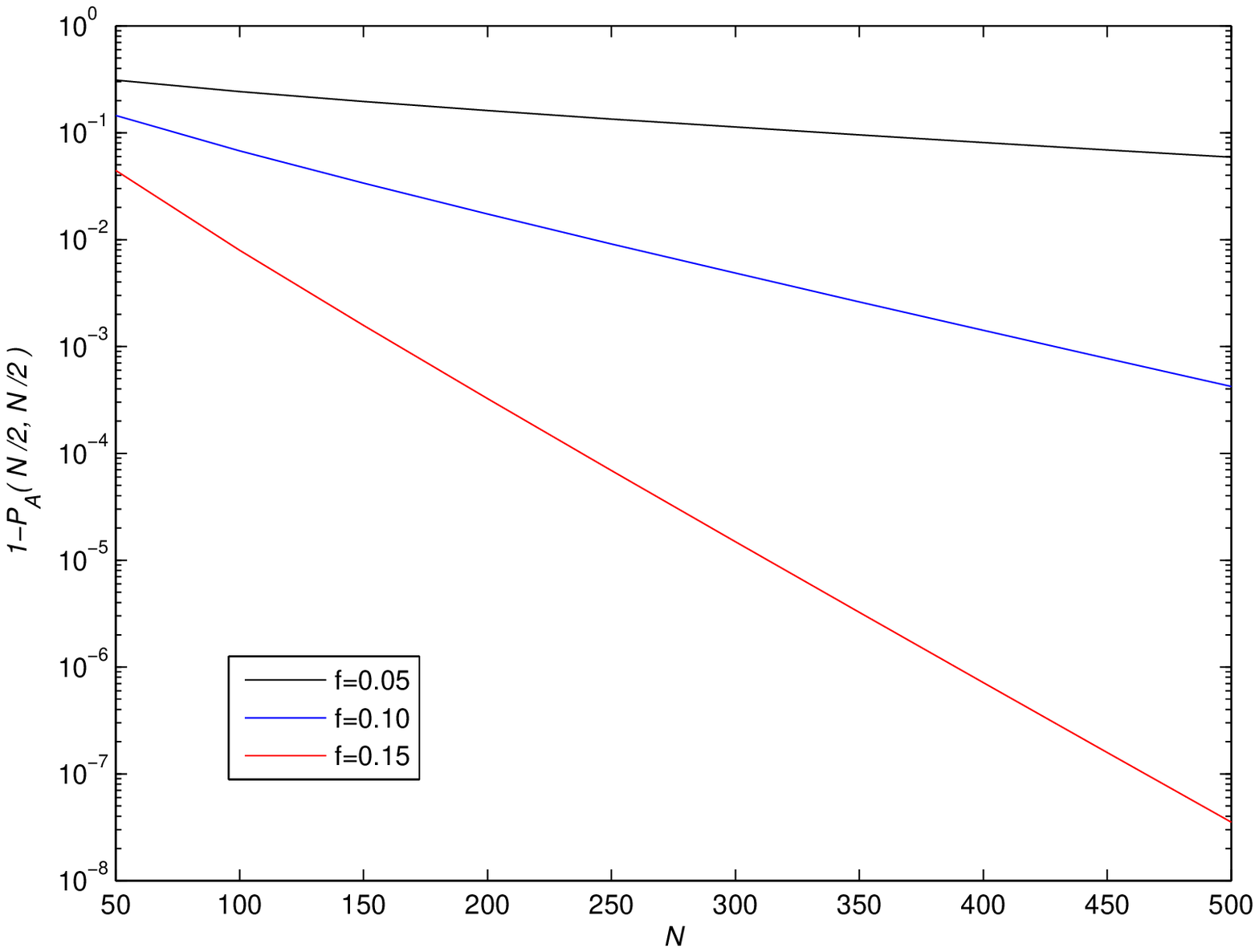}\\
  \caption{Probability of all B consensus $1-P_A$ starting from macrostate $(n_A,n_B)=(N/2,N/2)$ as a function of network size $N$ with different external influence level $f$'s.}
  \label{figure:8}
\end{figure}
\end{center}

Furthermore, we show in Fig.~\ref{figure:8} that the external
influence becomes more powerful in forcing the network to a desired
consensus state when network size $N$ grows larger. Starting from an
unbiased macrostate $(N/2,N/2)$, the probability of going to all B
consensus $1-P_A(N/2,N/2)$ (which is against of the central
influence) decays exponentially along with the network size $N$.
So it is reasonable to expect that in real social network for which network size $N$ is very large, a very slight biased central influence can strongly affect the social consensus.

\section{2-word Naming Game with committed agents}
A second method of speeding up consensus is by introducing an intrinsic
bias in the system, through a set of inflexible agents \cite{Galam_PhysA2007,Lucomm} promoting a designated opinion.
We refer to such individuals as {\it committed} agents \cite{Lucomm}.
Intuitively, the introduction of committed agents will break the
symmetry of the original NG and will facilitate global consensus to the
state adopted by the committed agents. In this section we provide asymptotic solutions of
2-word NG with committed agents

Suppose that the number of committed agents is $n_q$ and all
committed agents are in state $A$. The corresponding events in the
NG with committed agents and the associated RW transition
probabilities are summarized in Table~\ref{table:4}.
\begin{table}
\caption{Update events for the 2-word naming game with committed agents and the associated random walk transition probabilities}
\begin{center}
\begin{tabular}{c|c|c|c|l}\hline
speaker&listener & event &$\D \vec{n}(n_A,n_B)$  & probabilty \\ \hline
B or AB &A   &$A\go AB$ &(-1,0) & $P(A-)=(n_A-n_q) (N-n_A+n_B)/2N^2$\\
A or AB &AB  &$AB\go A$ &(1,0)  & $P(A+)=(N-n_A-n_B)(N+n_A-n_B)/ 2N^2$\\
A or AB &B   &$B\go AB$ &(0,-1) & $P(B-)=n_B (N+n_A-n_B)/ 2N^2$\\
B or AB &AB  &$AB\go B$ &(0,1)  & $P(B+)=(N-n_A-n_B)(N-n_A+n_B)/ 2N^2$\\
A, B or AB &A or B &unchanged &(0,0) & $P_0=(n_A+n_B)/2N+(n_A-n_B)^2/2N^2+n_q(N-n_A+n_B)/2N^2$ \\\hline
\end{tabular}
\end{center}
\label{table:4}
\end{table}
The equations $\vec{\tau}=Q\vec{\tau}+\vec{t}$ and $(I-Q)^T
u=\d_N^{\vec{n}}$ are derived in exactly the same way as in the
non-committed case.

\begin{center}
\begin{figure}[!htbp]
  \includegraphics[width=0.6\textwidth]{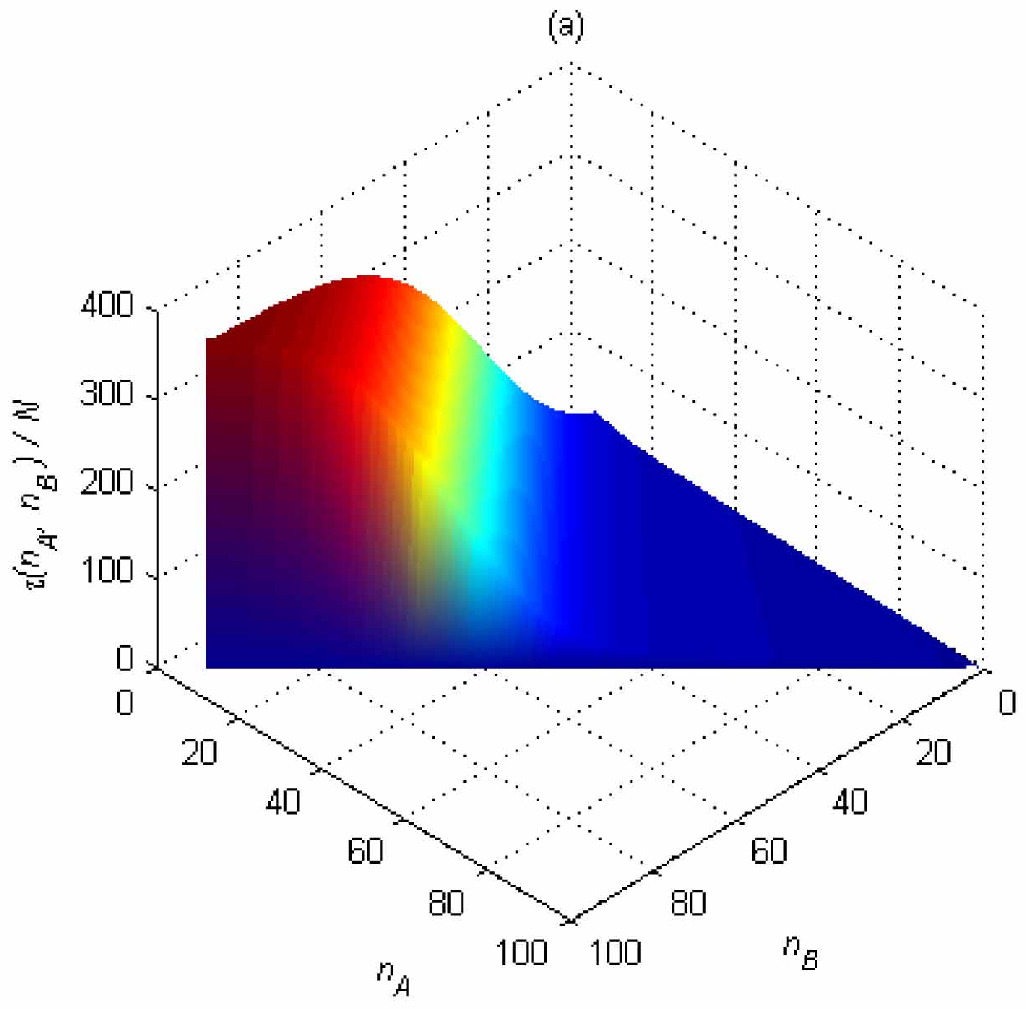}\\
  \includegraphics[width=0.6\textwidth]{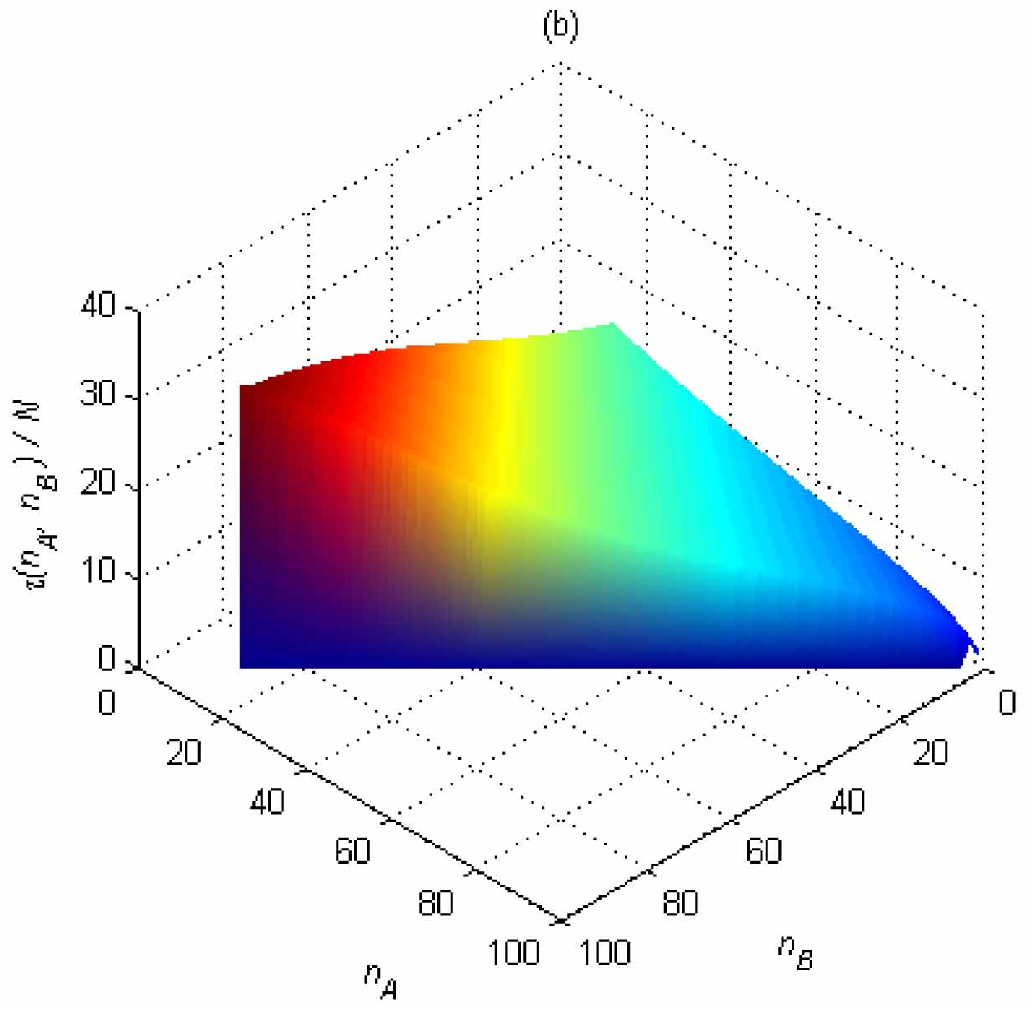}\\
  \caption{Expected normalized consensus time ($\tau(\vec{n})/N$) as a function of the initial macrostate $(n_A,n_B)$ on the
  complete graph with $N=100$ nodes.
  (a) when the fraction of committed agents is $q=0.06<q_c$;
  (b) when $q=0.12>q_c$.}
  \label{figure:9}
\end{figure}
\end{center}

From the dynamics of infinite systems with homogeneous mixing, one
can expect \cite{Galam_PhysA2007,Xie_2011}, that there exists a
critical value $q_c$ of the committed fraction $q = n_q/N$, above
which consensus times drop dramatically. More specifically, for
$q<q_c$ the phase space exhibits three fixed points: a
``meta-stable" one, dominated by individuals in state B, an stable
absorbing fixed point with all individuals in state A, and a
``saddle" point separating them. Initializing the system in a
configuration corresponding to macrostate $(n_A(0),n_B(0))=(n_q,
N-n_q)$ (a small number of committed agents embedded among Bs), the
system quickly relaxes to the meta-stable fixed point and stays
there for times exponentially large with the system size (i.e.,
forever in an infinite system). As $q$$\to$$q_c$, the meta-stable
fixed point and the saddle point merge, and become an unstable fixed
point. For $q>q_c$, regardless of the initial configuration, the
system quickly relaxes to the all-A absorbing fixed point.
Figure~\ref{figure:9} confirms the above scenario.
Figure~\ref{figure:9}(a) and (b) show the consensus times as a
function of the initial macrostate $(n_A,n_B)$ for $q<q_c$ and
$q>q_c$, respectively. In the former case, $q<q_c$, the expected
consensus time $\tau(n_A,n_B)$ is given by the equation
$\vec{\tau}=Q\vec{\tau}+\vec{t}$.
A fast drop-off in consensus time as a function of the initial
configuration, observed for $q<q_c$ [Fig.~\ref{figure:9}(a)],
indicates the presence of the saddle point. For $q>q_c$, there is
only one stable fixed point (the absorbing one), so the consensus
time starting from any initial configuration is short, including
initial configurations in the vicinity of the previously meta-stable
states [Fig.~\ref{figure:9}(b)].

\begin{center}
\begin{figure}[!htbp]
  \includegraphics[width=0.6\textwidth]{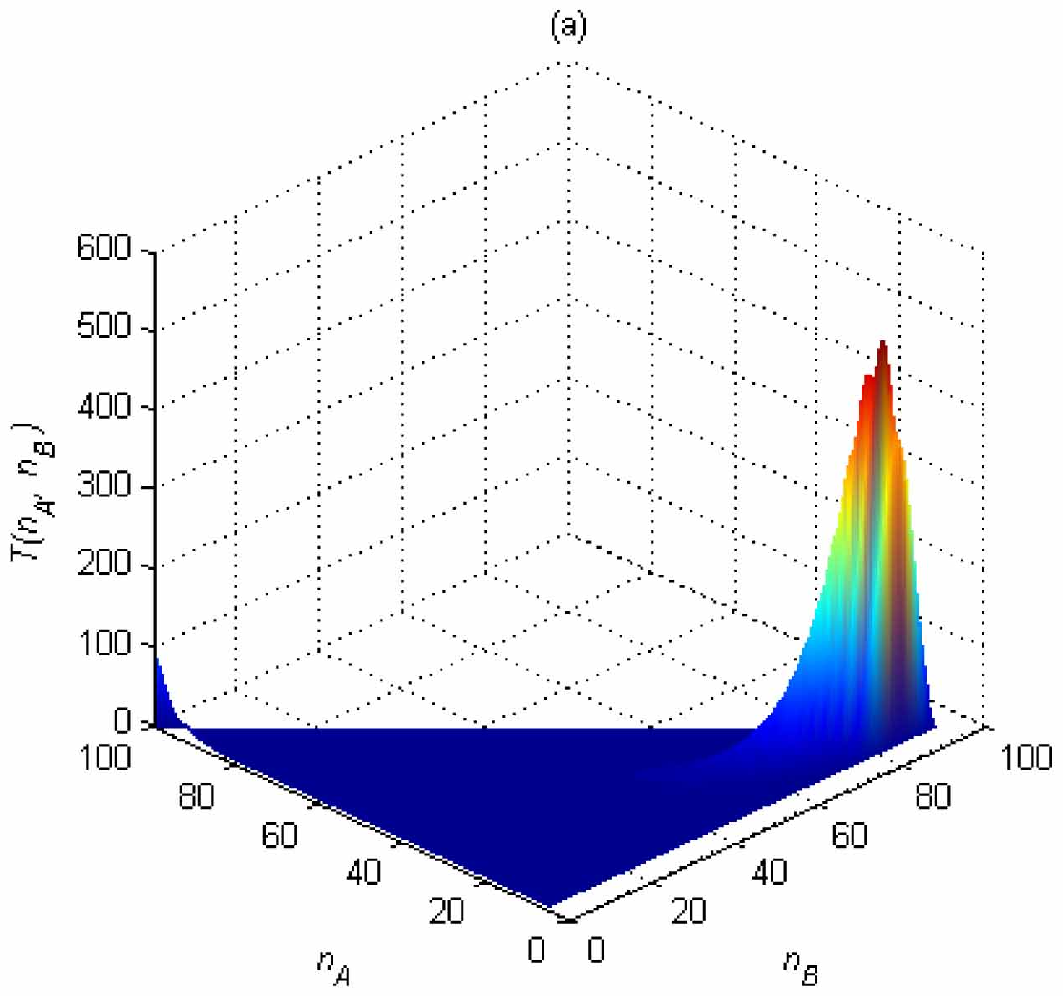}\\
  \includegraphics[width=0.6\textwidth]{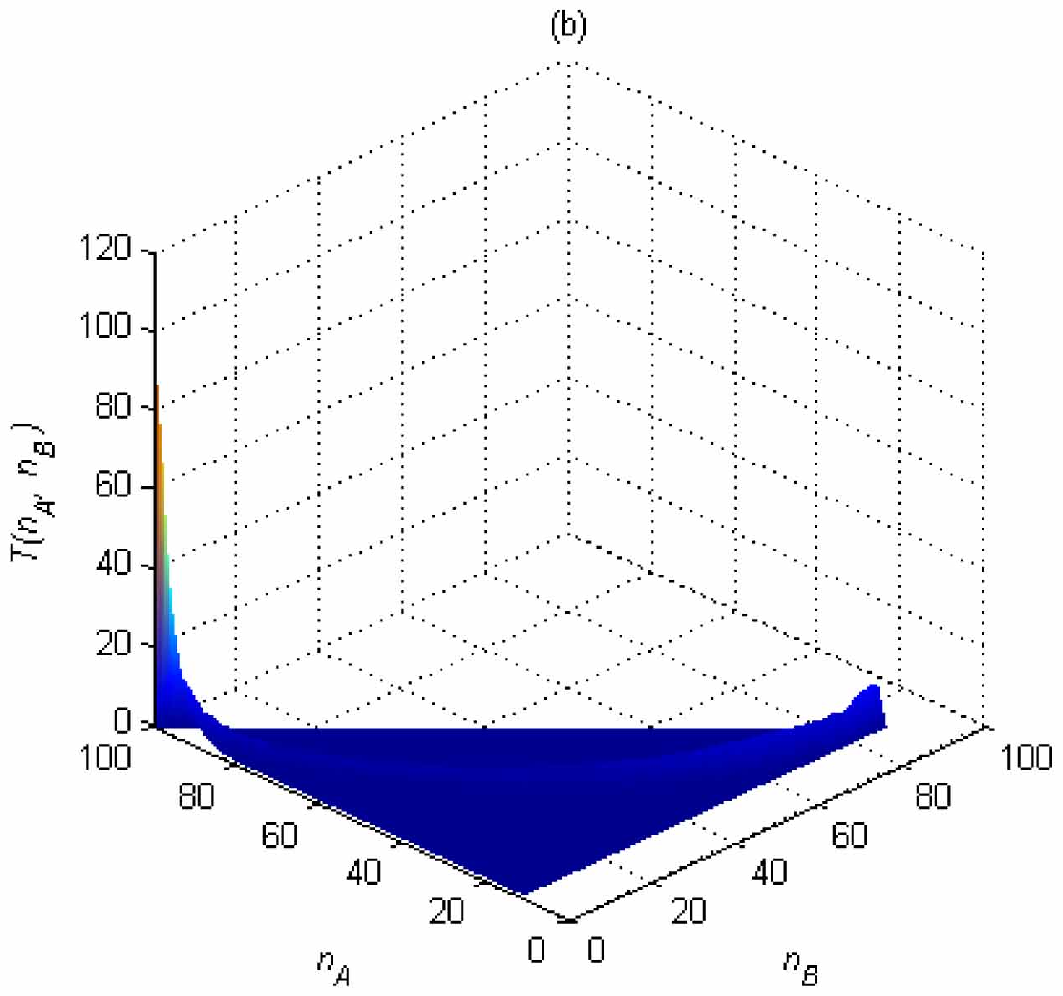}\\
  \caption{Expected time spent in each macrostate before consensus $T(n_A,n_B)$ on the complete graph
  with $N=100$ nodes, starting from the $(n_A(0),n_B(0))=(n_q,N-n_q)$ initial
  macrostate,
  (a) for $q=0.06<q_c$;
  (b) for $q=0.12>q_c$.}
  \label{figure:10}
\end{figure}
\end{center}

Figure \ref{figure:10}(a) and (b) present the expected number of
time steps spent before absorption in each macrostate $T(n_A,n_B)$
starting from the initial state $(n_q, N-n_q)$ for $q<q_c$ and
$q>q_c$, respectively. From Fig.~\ref{figure:10}(a) and (b),
according to the two peaks in each figure, the random walk before
absorption spend time mainly in two areas: one is near the
meta-stable state [close to the initial state $(n_q, N-n_q)$], the
other one is around the consensus state $(N,0)$.  When $q<q_c$, the
peak around the meta-stable state is dominant in the total consensus
time, while for $q>q_c$, it can be ignored compared to the peak
around consensus.

\begin{center}
\begin{figure}[!htbp]
  \includegraphics[width=0.7\textwidth]{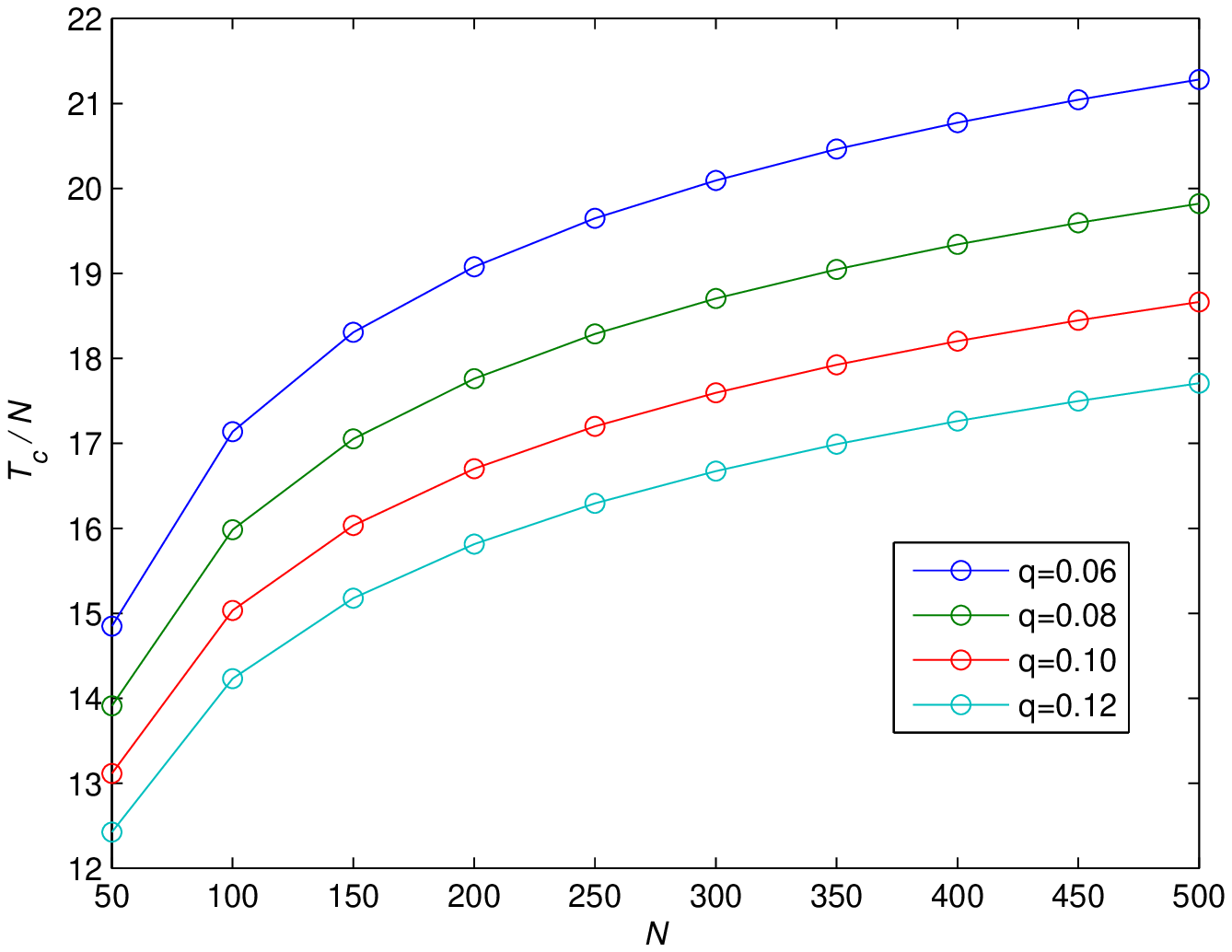}\\
  \caption{Normalized time spent near the consensus state before consensus as a function of network size $N$ for
  different fraction of committed agents $q$, including cases for both $q<q_c$ and $q>q_c$.}
  \label{figure:11}
\end{figure}
\end{center}

\begin{center}
\begin{figure}[!htbp]
  \includegraphics[width=0.7\textwidth]{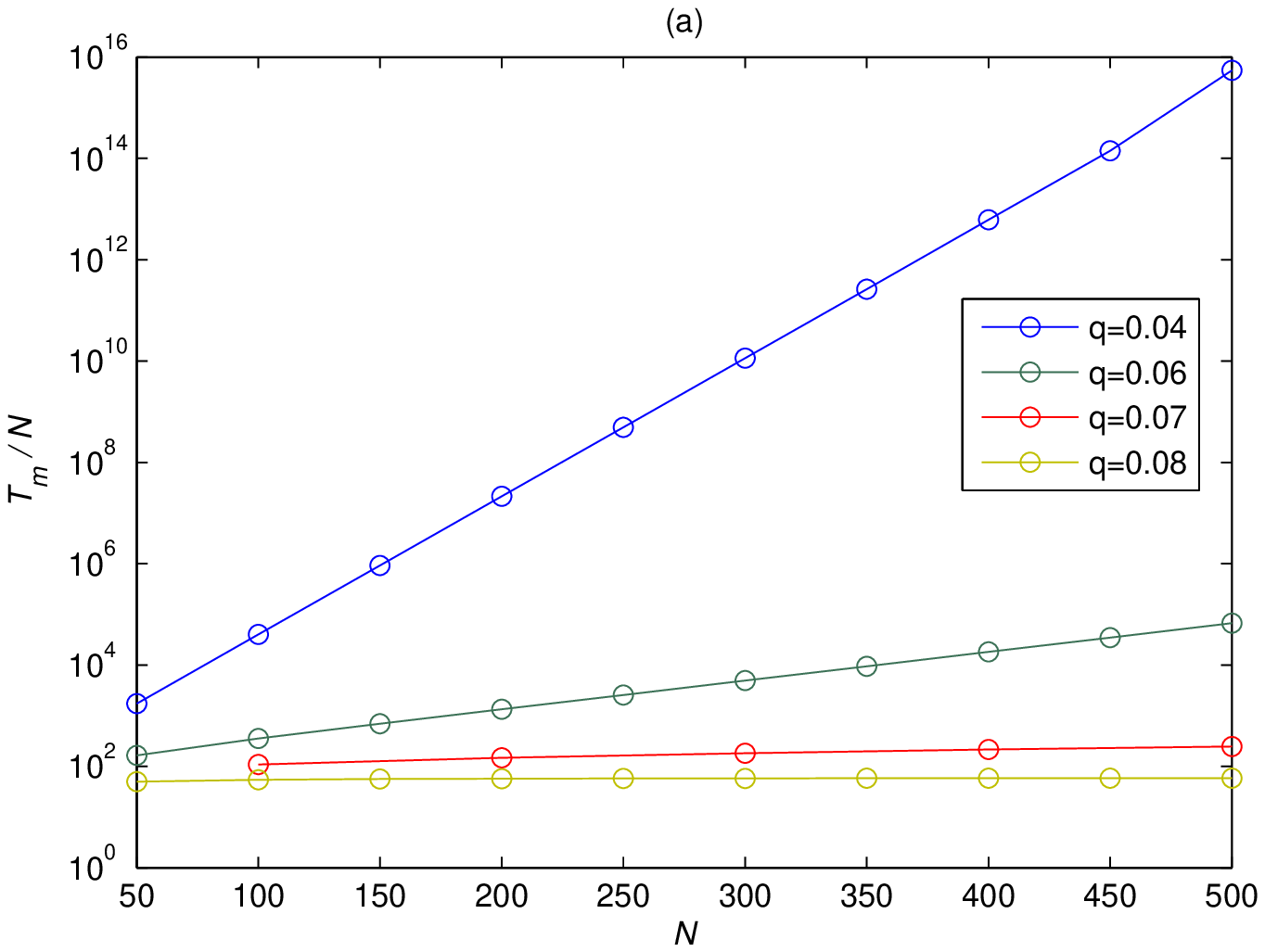}\\
  \includegraphics[width=0.7\textwidth]{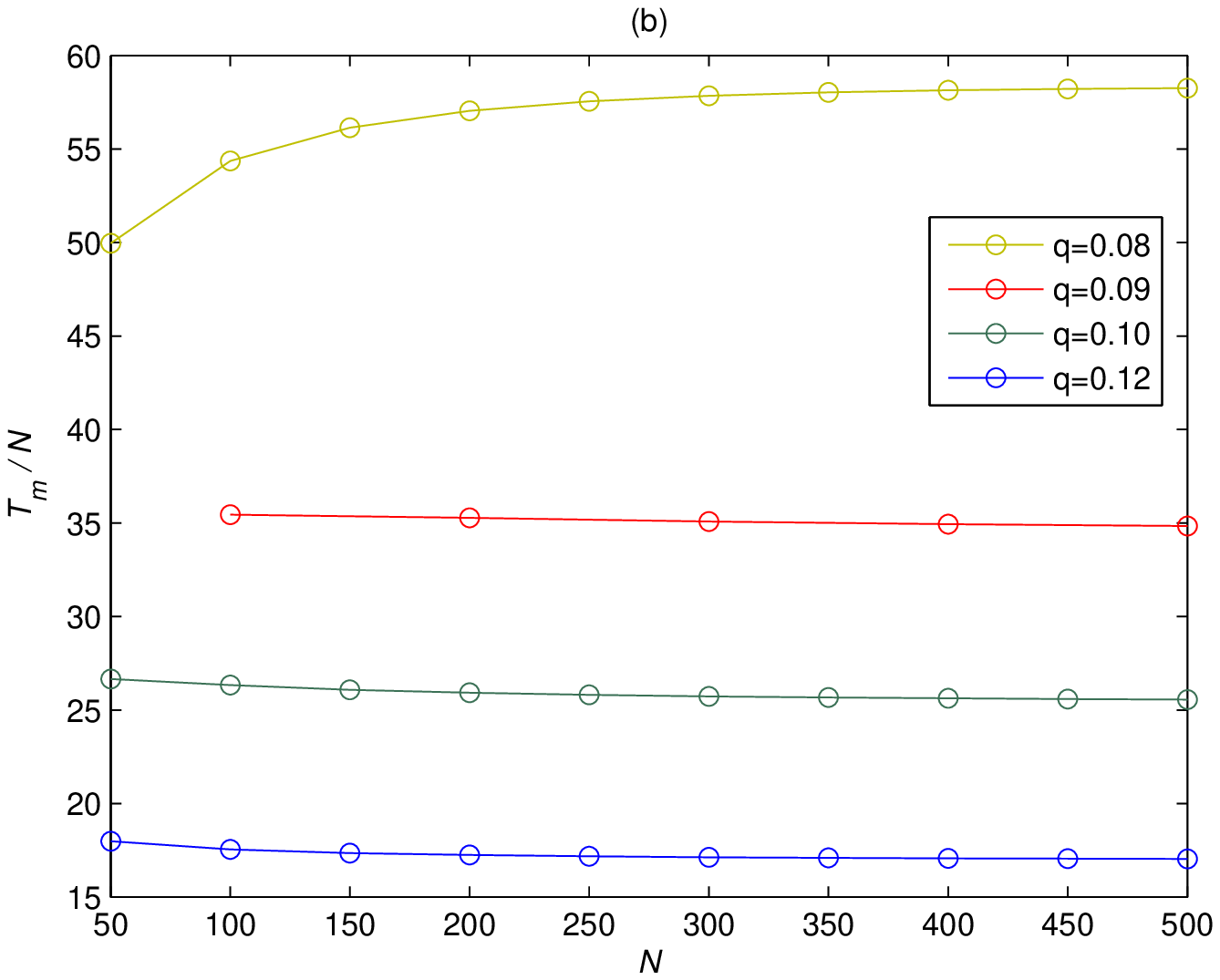}\\
  \caption{Normalized time spent near the meta-stable state as a function of network size $N$
  for different fraction of committed agents $q$, (a) for $q<q_c$; (b) for $q>q_c$. The behavior for $q=0.08$ is shown in both
  (a) and (b), corresponding to our rough estimate of the critical fraction of committed agents, $q_c\approx0.08\pm0.01$.}
  \label{figure:12}
\end{figure}
\end{center}

We sum $T(n_A,n_B)$ over these two areas separately (since the peak
in Fig.~\ref{figure:10} is very concentrated, the domain of
summation does not matter very much), and define the time spent near
consensus state as $T_c=\sum_{n_A>n_B}T(n_A,n_B)$ and that near the
meta-stable state as $T_m=\sum_{n_A>n_B}T(n_A,n_B)$.
Figure~\ref{figure:11} shows that the normalized time spent near
consensus $T_c/N$ has the same order $O(\ln(N))$ when $N$ grows as
demonstrated in the non-committed case. We conclude that for
different $q$'s regardless whether it is less or greater than the
critical $q_c$, the peaks around the consensus state have roughly
the same scale and are about twice the height of the corresponding
peak in the 2-word NG without committed agents shown in
Fig.~\ref{figure:4}. Figure~\ref{figure:12} shows that the
normalized time that the random walk is stuck in the vicinity of the meta-stable
state, $T_m/N$, grows exponentially with $N$ for $q<q_c$, while it
decreases weakly with $N$ for $q>q_c$. The crossover and departure
from these drastically different scaling behaviors appears at around
$q=0.08$, hence our rough estimate for the critical fraction of
committed agents is $q_c\approx0.08\pm0.01$. A detailed finite-size
analysis of this crossover behavior should be performed to extract
$q_c$ in the infinite system-size limit. Since for the consensus
time we approximately have $\tau\approx T_m+T_c$, the above findings
imply that $\tau/N\sim O(e^{cN})$ (where $c$ is a constant) for
$q<q_c$, while $\tau/N\sim O(\ln(N))$ for $q>q_c$.

\section{Summary}

We studied influencing and consensus formation, in particular, the
asymptotic consensus times, in stylized individual-based models for
social dynamics on the complete graph. We accomplished this by a
coarse-graining approach (lumping microscopic variables into
macrostates) resulting in an associated random walk. We then
analyzed first-passage times (corresponding to consensus) and times
spent in each macro-state of the random-walk model. The method
yields asymptotically exact consensus times for large but finite
complete graphs of size $N$. Direct individual-based simulations can
become time consuming for large systems and prohibitive even for
moderately-sized systems when the system initially is in a
meta-stable configuration, as the escape time can increase
exponentially with the system size. The method presented here
provides an alternative way to obtain the asymptotic behavior of
consensus times, including the cases associated with extremely slow
meta-stable escapes.

After testing this framework on spontaneous opinion formation in two
known models, we applied it to two scenarios for social influencing
in a variation of the 2-word naming game. First, we considered the
case when individuals are exposed to a global external field (or
central influence). We found that the external field dominates the
consensus in the large network-size limit. Second, we investigated
the impact of committed individuals with a fixed designated opinion
(i.e., individuals who can influence others but themselves are immune to influence).
In the latter case, we found the existence of a tipping point,
associated with the disappearance of the meta-stable state in the
opinion space: When the fraction of committed nodes is below a
critical value, consensus times increase exponentially with system
size; on the other hand when the fraction of committed nodes is
above this threshold value (tipping point) the system is quickly
driven to consensus with weak system size dependence.

While the method and the results presented here are applicable to
the complete graph, consensus times often exhibit the same
asymptotic scaling with the system size in large homogeneous sparse
random networks
\cite{Castelano_PRE2005,DallAsta_PRE2006,Lu_PRE2008}, hence our
results yield some insight how ordering and consensus can evolve in
realistic social networks. In particular, one can better understand
and predict timescales associated with reaching consensus in social
networks.

\section*{Acknowledgements}
This work was supported in part by the Army Research Laboratory
under Cooperative Agreement Number W911NF-09-2-0053, by the Army
Research Office Grant No. W911NF-09-1-0254, and by the Office of
Naval Research Grant No. N00014-09-1-0607. The views and conclusions
contained in this document are those of the authors and should not
be interpreted as representing the official policies, either
expressed or implied, of the Army Research Laboratory or the U.S.
Government.

\end{document}